\documentclass{article}
\usepackage{amssymb}
\usepackage{amsmath}

\input{tcilatex}
\begin{document}

\begin{center}
459SolvDynSystPlaneWithPolynInteract190122

(Submitted to:\ J. Phys.\ Conf. Ser., 22.01.2019)

\bigskip

{\LARGE Some Algebraically Solvable Two-Dimensional Dynamical Systems with
Polynomial Interactions}

\bigskip

\textbf{Francesco Calogero}$^{a,b,1}${\LARGE \ {\large and} }\textbf{Farrin
Payandeh}$^{a,c,2}$

$^{a}$ Physics Department, University of Rome "La Sapienza", Rome, Italy

$^{b}$ INFN, Sezione di Roma 1

$^{c}$ Department of Physics, Payame Noor University (PNU), PO BOX
19395-3697 Tehran, Iran

$^{1}$ francesco.calogero@roma1.infn.it, francesco.calogero@uniroma1.it

$^{2}$ f\_payandeh@pnu.ac.ir, farrinpayandeh@yahoo.com

\bigskip

\textit{Abstract}
\end{center}

We tersely review a recently introduced technique to identify systems of two
nonlinearly-coupled Ordinary Differential Equations (ODEs) \textit{solvable
by algebraic operations}; and we report some specific examples of this kind,
namely systems of $2$ first-order ODEs with polynomial right-hand sides,%
\begin{equation*}
\dot{x}_{n}=P^{\left( n\right) }\left( x_{1},x_{2}\right) ~,~~~n=1,2~,
\end{equation*}%
satisfied by the $2$ (possibly \textit{complex}) dependent variables $%
x_{n}\equiv x_{n}\left( t\right) $. Here $P^{\left( n\right) }\left(
x_{1},x_{2}\right) $ indicates some specific \textit{polynomial}. These
examples are analogous, but different, from those previously reported.

\bigskip

\section{Introduction}

A technique to identify dynamical systems characterized by systems of
\textit{nonlinearly-coupled} Ordinary Differential Equations (ODEs) \textit{%
solvable by algebraic operations} is based on the relations among the $N$
\textit{coefficients} $y_{m}\left( t\right) $ and the $N$ \textit{zeros} $%
x_{n}\left( t\right) $ of a time-dependent (monic) polynomial of degree $N$
in the (complex) variable $z$:%
\begin{equation}
p_{N}\left( z;t\right) =z^{N}+\sum_{m=1}^{N}\left[ y_{m}\left( t\right)
~z^{N-m}\right] =\dprod\limits_{n=1}^{N}\left[ z-x_{n}\left( t\right) \right]
~.  \label{PolN}
\end{equation}%
The basic idea is to consider "simple" time evolutions of the $N$ \textit{%
coefficients} $y_{m}\left( t\right) $---possibly \textit{explicitly solvable}
evolutions, maybe featuring remarkable properties such as \textit{isochrony}
(implying that \textit{all} the $N$ coefficients $y_{m}\left( t\right) $ are
periodic in $t$ with the same fixed period independent of the initial data);
and to then consider the corresponding time evolutions of the $N$ zeros $%
x_{n}\left( t\right) $, which are generally characterized by "more
nonlinear" equations of motions, the solutions of which are then obtainable
via the \textit{algebraic} operation of computing the $N$ zeros $x_{n}\left(
t\right) $ of the polynomial $p_{N}\left( z;t\right) $ defined in terms of
its $N$ \textit{coefficients} $y_{m}\left( t\right) .$ This generally
entails that the time evolution of the \textit{zeros} $x_{n}\left( t\right) $
\textit{inherits} properties (such as \textit{isochrony}) of the evolution
of the coefficients $y_{m}\left( t\right) $. This idea has a long history
\cite{C1978}, and it has produced many developments, see for instance \cite%
{C2001} \cite{C2008} \cite{C2018a} \cite{C2018b} \cite{GS2005} and
references therein.

An additional recent development has explored the modifications of the
approach outlined above which emerge if, rather than considering \textit{%
generic} time-dependent polynomials, one focusses on \textit{special}
time-dependent polynomials featuring (for all time) \textit{multiple zeros}:
in particular, as a first step in this direction, results have been reported
for the case of a polynomial featuring \textit{one double zero }\cite{BC2018}%
, then some significant progress has been made on the case of a polynomial
featuring \textit{one zero of arbitrary multiplicity} \cite{B2018}, and
finally the most general case has been treated of a polynomial $p_{M}\left(
z;t\right) $ of degree $M$ featuring an \textit{arbitrary} number $N$ of
\textit{zeros} each of them of \textit{arbitrary multiplicity} $\mu _{n}$
\cite{CP2019a}:
\begin{subequations}
\label{pM}
\begin{equation}
p_{M}\left( z;t\right) =z^{N}+\sum_{m=1}^{M}\left[ y_{m}\left( t\right)
~z^{M-m}\right] =\dprod\limits_{n=1}^{N}\left\{ \left[ z-x_{n}\left(
t\right) \right] ^{\mu _{n}}\right\} ~,  \label{PM}
\end{equation}%
of course with the $N$ positive integers $\mu _{n}$ related to the positive
integer parameters $M$ and $N$ by the relation%
\begin{equation}
M=\sum_{n=1}^{N}\left( \mu _{n}\right) ~.  \label{M}
\end{equation}

A second twist of this development restricted attention to the $N=2$
case---i. e., polynomials (see (\ref{pM})) featuring only $2$ \textit{zeros}
\cite{CP2019b} \cite{CP2019a}. This stringent limitation has opened the way
to the identification of \textit{algebraically solvable two-dimensional
dynamical systems with polynomial interactions}, namely systems
characterized by the following system of two nonlinearly coupled ODEs:
\end{subequations}
\begin{equation}
\dot{x}_{n}=P^{\left( n\right) }\left( x_{1},x_{2}\right) ~,~~~n=1,2~.
\label{PolModels}
\end{equation}

\textbf{Notation 1.1}. Here and hereafter superimposed dots denote
differentiations with respect to the dependent variable $t$ ("time"), $%
x_{n}\equiv x_{n}\left( t\right) $ are the $2$ (generally \textit{complex})
dependent variables, and $P^{\left( n\right) }\left( x_{1},x_{2}\right) $ is
generally a polynomial (or "almost a polynomial": see below) in these $2$
dependent variables $x_{1},~x_{2}$. Parameters such $a,$ $b,$ $c,$ $\alpha ,$
$\beta ,$ $\gamma $ (possibly equipped with subscripts) are
time-independent; they can be arbitrarily assigned (up to explicitly
indicated relations among them). Note that often, above and hereafter, the $%
t $-dependence of variables is \textit{not} explicitly indicated (when this
is unlikely to generate any misunderstanding). $\square $

A few such systems have been identified and tersely discussed in \cite%
{CP2019b} by taking as point of departure the case of a polynomial such as (%
\ref{PM}) with $M=3$, and several such cases have been treated in \cite%
{CP2019a} by focussing on the case with $M=4$. In the present paper---after
a terse reminder of the methodology to obtain these results---we focus again
on the $M=3$ case, reporting several solvable models of type (\ref{PolModels}%
) not previously identified. Analogous treatments of cases with $M>4$ shall
be performed by ourselves or by others in the future.

In the following \textbf{Section 2} we tersely review our notation and some
results which are basic for the following treatment. In \textbf{Section 3}\
and its subsections several specific examples are reported (with some
related results confined to \textbf{Appendix A}). The last \textbf{Section 4
}outlines possible future developments.

\bigskip

\section{Preliminaries}

A basic tool of our treatment are the following definitions of the $3$
variables $y_{m}\equiv y_{m}\left( t\right) ,$ $m=1,2,3$ in terms of the $2$
variables $x_{n}\equiv x_{n}\left( t\right) ,$ $n=1,2$:
\begin{subequations}
\begin{equation}
y_{1}=-\left( 2x_{1}+x_{2}\right) ~,~~~y_{2}=x_{1}\left( x_{1}+2x_{2}\right)
~,~~\ y_{3}=-\left( x_{1}\right) ^{2}x_{2}~;  \label{ym}
\end{equation}%
they of course imply (see (\ref{PM})) that $x_{1}$ and $x_{2}$ are the $2$
\textit{zeros}---with respective multiplicities $2$ and $1$---of the (monic)
third-degree polynomial $p_{3}\left( z;t\right) $ with coefficients $y_{1},$
$y_{2},$ $y_{3}$:%
\begin{equation}
p_{3}\left( z;t\right) =z^{3}+\sum_{m=1}^{3}\left[ y_{m}\left( t\right)
~z^{3-m}\right] =\left[ z-x_{1}\left( t\right) \right] ^{2}~\left[
z-x_{2}\left( t\right) \right] ~.  \label{p3zt}
\end{equation}%
Note that this entails that $x_{1}$ and $x_{2}$ can be obtained---in terms
of $y_{1}$ and $y_{2}$, or $y_{1}$ and $y_{3}$, or $y_{2}$ and $y_{3}$---by
solving the following algebraic equations:
\end{subequations}
\begin{subequations}
\label{xyy}
\begin{equation}
3\left( x_{1}\right) ^{2}+2y_{1}x_{1}+y_{2}=0~,~~~x_{2}=-\left(
2x_{1}+y_{1}\right) ~,  \label{xy1y2}
\end{equation}%
or%
\begin{equation}
2\left( x_{1}\right) ^{3}+y_{1}\left( x_{1}\right)
^{2}+y_{3}=0~,~~~x_{2}=-\left( 2x_{1}+y_{1}\right) ~,  \label{xy1y3}
\end{equation}%
or%
\begin{equation}
\left( x_{1}\right) ^{3}-y_{2}x_{1}-2y_{3}=0~,~~~x_{2}=\frac{-\left(
x_{1}\right) ^{2}+y_{2}}{2x_{1}}~.  \label{xy2y3}
\end{equation}

It can moreover be easily shown---or see \cite{BC2018} or \cite{CP2019a} or
\cite{CP2019b}---that these formulas imply the following differential
relations:
\end{subequations}
\begin{subequations}
\label{xndotyy}
\begin{equation}
\dot{x}_{1}=-\frac{2x_{1}\dot{y}_{1}+\dot{y}_{2}}{2\left( x_{1}-x_{2}\right)
}~,~~~\dot{x}_{2}=\frac{\left( x_{1}+x_{2}\right) \dot{y}_{1}+\dot{y}_{2}}{%
x_{1}-x_{2}}~,  \label{xndoty1y2}
\end{equation}%
\begin{equation}
\dot{x}_{1}=\frac{-\left( x_{1}\right) ^{2}\dot{y}_{1}+\dot{y}_{3}}{%
2x_{1}\left( x_{1}-x_{2}\right) }~,~~~\dot{x}_{2}=\frac{x_{1}x_{2}\dot{y}%
_{1}-\dot{y}_{3}}{x_{1}\left( x_{1}-x_{2}\right) }~,  \label{xndoty1y3}
\end{equation}%
\begin{equation}
\dot{x}_{1}=\frac{x_{1}\dot{y}_{2}+2\dot{y}_{3}}{2x_{1}\left(
x_{1}-x_{2}\right) }~,~~~\dot{x}_{2}=-\frac{x_{1}x_{2}\dot{y}_{2}+\left(
x_{1}+x_{2}\right) \dot{y}_{3}}{\left( x_{1}\right) ^{2}\left(
x_{1}-x_{2}\right) }~.  \label{xndoty2y3}
\end{equation}

It is plain from these formulas that if the two variables $y_{1}\equiv
y_{1}\left( t\right) $ and $y_{2}\equiv y_{2}\left( t\right) $, or $%
y_{1}\equiv y_{1}\left( t\right) $ and $y_{3}\equiv y_{3}\left( t\right) $,
or $y_{2}\equiv y_{2}\left( t\right) $ and $y_{3}\equiv y_{3}\left( t\right)
$ satisfy an \textit{algebraically solvable} system of $2$ first-order ODEs,
say
\end{subequations}
\begin{equation}
\dot{y}_{m_{1}}=f_{m_{1}}\left( y_{m_{1}},y_{m_{2}}\right) ~,~~~\dot{y}%
_{m_{2}}=f_{m_{2}}\left( y_{m_{1}},y_{m_{2}}\right)  \label{ydotf}
\end{equation}%
with $m_{1}=1,$ $m_{2}=2$ or $m_{1}=1,$ $m_{2}=3$ or $m_{1}=2,$ $m_{2}=3$,
then the corresponding system of $2$---generally nonlinearly-coupled,
first-order---ODEs satisfied by $x_{1}\equiv x_{1}\left( t\right) $ and $%
x_{2}\equiv x_{2}\left( t\right) $ is as well \textit{algebraically solvable}%
. Moreover, if the functions $f_{m_{1}}\left( y_{m_{1}},y_{m_{2}}\right) $
and $f_{m_{2}}\left( y_{m_{1}},y_{m_{2}}\right) $ are \textit{appropriately
chosen}, then the right-hand sides of the ODEs (\ref{xndotyy}) become
\textit{polynomial} (or "\textit{almost polynomial}": see below). In the
following \textbf{Section 3} we report several \textit{algebraically solvable%
} systems satisfied by the $2$ dependent variables $x_{1}\equiv x_{1}\left(
t\right) $ and $x_{2}\equiv x_{2}\left( t\right) $ obtained in this manner,
i. e. by first replacing in the relevant equations (\ref{xndotyy}) the
expressions of the time-derivatives of the relevant $2$ variables $y_{m}$
via (\ref{ydotf}) and subsequently replacing the expressions of these $2$
variables $y_{m}$ in terms of the $2$ variables $x_{n}$ via the relevant
equations (\ref{ym}); of course with the functions $f_{m_{1}}\left(
y_{m_{1}},y_{m_{2}}\right) $ and $f_{m_{2}}\left( y_{m_{1}},y_{m_{2}}\right)
$ \textit{appropriately chosen}.

\bigskip

\section{Results}

Our first step is to identify $3$ \textit{solvable} systems of $2$ evolution
equations satisfied by the dependent variables $y_{m_{1}}\equiv
y_{m_{1}}\left( t\right) $ and $y_{m_{2}}\equiv y_{m_{2}}\left( t\right) $.
Their solvability is discussed in \textbf{Appendix A }of Ref. \cite{CP2019b}
and tersely reviewed below in \textbf{Appendix\ A}.

The first of these $3$ systems---hereafter identified as \textbf{A1} (see
\textbf{Appendix A})---is characterized by the following $2$ \textit{%
uncoupled} ODEs:%
\begin{equation}
\dot{y}_{m_{1}}=\sum_{\ell =0}^{L}\left[ \alpha _{\ell }\left(
y_{m_{1}}\right) ^{\ell m_{2}+1}\right] ~,~~~\dot{y}_{m_{2}}=\sum_{\ell
=0}^{L}\left[ \beta _{\ell }\left( y_{m_{2}}\right) ^{\ell m_{1}+1}\right] ~.
\label{A1}
\end{equation}%
\

The second of these $3$ systems---hereafter identified as \textbf{A2} (see
\textbf{Appendix A})---is characterized by the following $2$ \textit{coupled}
ODEs:%
\begin{equation}
\dot{y}_{m_{1}}=\sum_{\ell =0}^{L}\left[ \alpha _{\ell }\left(
y_{m_{1}}\right) ^{\ell }\right] ~,~~~\dot{y}_{m_{2}}=y_{m_{2}}\sum_{\ell
=1}^{L}\left[ \beta _{\ell }\left( y_{m_{1}}\right) ^{\ell -1}\right]
+\sum_{\ell =0}^{L}\left[ \gamma _{\ell }\left( y_{m_{1}}\right) ^{\ell
-1+\left( m_{2}/m_{1}\right) }\right] ~.  \label{A2}
\end{equation}

The third of these $3$ systems---hereafter identified as \textbf{A3} (see
\textbf{Appendix A})---is characterized by the following $2$ coupled ODEs:%
\begin{eqnarray}
\dot{y}_{m_{1}} &=&\alpha _{0}+\alpha _{1}y_{m_{2}}~,~~~\dot{y}%
_{m_{2}}=\beta _{0}\left( y_{m_{1}}\right) ^{-1+m}+\beta _{1}\left(
y_{m_{1}}\right) ^{-1+2m}~,  \notag \\
m &=&m_{2}/m_{1}~,~~~m_{1}=1~,~~~m_{2}=2,3~.  \label{A3}
\end{eqnarray}

Our next step is to list in the following $11$ subsections $11$ \textit{%
algebraically solvable} systems of $2$ nonlinearly-coupled ODEs with \textit{%
polynomial} (or "\textit{almost polynomial}": see below) right-hand sides
(see (\ref{PolModels})); in each case we identify the corresponding \textit{%
appropriately chosen algebraically solvable} system of ODEs---see above and
\textbf{Appendix A}---satisfied by the corresponding functions $%
y_{m_{1}}\left( t\right) $ and $y_{m_{2}}\left( t\right) $. But note that
the algebraically solvable systems thus identified are only $9$, because $2$
pairs of the systems identified below---although arrived at
differently---are in fact \textit{identical} (a phenomenon already noted in
\cite{CP2019a}).

\bigskip

\subsection{Models of type \textbf{A1}}

In the $6$ cases listed in this \textbf{Subsection} the variables $%
y_{m_{1}}\left( t\right) $ and $y_{m_{2}}\left( t\right) $ are supposed to
satisfy the system \textbf{A1} of $2$ uncoupled ODEs (see (\ref{A1}) and
\textbf{Appendix A}), with the indicated assignments of the various
parameters.

\textbf{Model A1.1}: $m_{1}=1,~m_{2}=2;$ $L=1;$ $\alpha _{0}=a,$~$\alpha
_{1}=b;~\beta _{0}=2a,~\beta _{1}=6b;$%
\begin{eqnarray}
\dot{x}_{1} &=&ax_{1}+bx_{1}\left[ 5\left( x_{1}\right)
^{2}+5x_{1}x_{2}-\left( x_{2}\right) ^{2}\right] ~,  \notag \\
\dot{x}_{2} &=&ax_{2}-b\left[ 2\left( x_{1}\right) ^{3}-2\left( x_{1}\right)
^{2}x_{2}-8x_{1}\left( x_{2}\right) ^{2}-\left( x_{2}\right) ^{3}\right] ~.
\label{A11}
\end{eqnarray}

\textbf{Model A1.2}: $m_{1}=1,~m_{2}=3;$ $L=1;$ $\alpha _{0}=a,$~$\alpha
_{1}=-2b;~\beta _{0}=3a,~\beta _{1}=-162b;$%
\begin{eqnarray}
\dot{x}_{1} &=&x_{1}\left\{ a+b\left[ 16\left( x_{1}\right) ^{3}+48\left(
x_{1}\right) ^{2}x_{2}-9x_{1}\left( x_{2}\right) ^{2}-\left( x_{2}\right)
^{3}\right] \right\} ~,  \notag \\
\dot{x}_{2} &=&x_{2}\left\{ a-2b\left[ 16\left( x_{1}\right) ^{3}-33\left(
x_{1}\right) ^{2}x_{2}-9x_{1}\left( x_{2}\right) ^{2}-\left( x_{2}\right)
^{3}\right] \right\} ~.  \label{A12}
\end{eqnarray}

\textbf{Model A1.3}: $m_{1}=2,~m_{2}=3;$ $L=1;$ $\alpha _{0}=2a,$~$\alpha
_{1}=2b;~\beta _{0}=3a,~\beta _{1}=81b;$%
\begin{eqnarray}
\dot{x}_{1} &=&x_{1}\left\{ a+b\left( x_{1}\right) ^{3}\left[ \left(
x_{1}\right) ^{3}+9\left( x_{1}\right) ^{2}x_{2}+33x_{1}\left( x_{2}\right)
^{2}-16\left( x_{2}\right) ^{3}\right] \right\} ~,  \notag \\
\dot{x}_{2} &=&x_{2}\left\{ a-b\left( x_{1}\right) ^{3}\left[ 2\left(
x_{1}\right) ^{3}+18\left( x_{1}\right) ^{2}x_{2}-15x_{1}\left( x_{2}\right)
^{2}-32\left( x_{2}\right) ^{3}\right] \right\} ~.  \label{A13}
\end{eqnarray}

\bigskip

\subsection{Models of type \textbf{A2}}

In the cases listed in this \textbf{Subsection} the variables $%
y_{m_{1}}\left( t\right) $ and $y_{m_{2}}\left( t\right) $ are supposed to
satisfy the system \textbf{A2} of $2$ coupled ODEs (see (\ref{A2}) and
\textbf{Appendix A}), with the indicated assignments of the various
parameters.

\textbf{Model A2.1}: $m_{1}=1,~m_{2}=2;$ $L=3;$ $\alpha _{0}=3b_{0},~\alpha
_{1}=a_{0}-3b_{1},~\alpha _{2}=-a_{1}+3b_{2},~\alpha _{3}=a_{2}-3b_{3};$ $%
\beta _{\ell }=\left( -1\right) ^{\ell -1}2a_{\ell -1},~\ell =1,2,3;~\gamma
_{\ell }=\left( -1\right) ^{\ell }2b_{\ell },~\ell =0,1,2,3;$%
\begin{eqnarray}
\dot{x}_{n} &=&x_{n}\left( a_{0}+a_{1}X+a_{2}X^{2}\right) -\left[
b_{0}+b_{1}X+b_{2}X^{2}+b_{3}X^{3}\right] ~,  \notag \\
n &=&1,2;~X=2x_{1}+x_{2}~.  \label{A21}
\end{eqnarray}

\textbf{Model A2.2}: $m_{1}=1,~m_{2}=2;$ $L=3;$ $\alpha _{0}=0,~\alpha
_{\ell }=c_{\ell };~\beta _{\ell }=2c_{\ell };~\ell =1,2,3~;$%
\begin{equation}
\dot{x}_{n}=x_{n}\left( c_{1}+c_{2}X+c_{3}X^{2}\right)
~,~~~n=1,2,~~~X=x_{1}\left( x_{1}+2x_{2}\right) ~.  \label{A22}
\end{equation}

\textbf{Model A2.3}: $m_{1}=1,~m_{2}=3;$ $L=3;$ $\alpha _{0}=9b_{0},$~$%
\alpha _{\ell }=\left( -1\right) ^{\ell -1}\left( a_{0}-9b_{\ell }\right)
,~\beta _{\ell }=\left( -1\right) ^{\ell -1}3a_{\ell -1},$ $\ell =1,2,3;$ $%
\gamma _{\ell }=\left( -1\right) ^{\ell }b_{\ell },~\ell =0,1,2,3.$%
\begin{eqnarray}
\dot{x}_{1} &=&x_{1}\left( a_{0}+a_{1}X+a_{2}X^{2}\right) -\left( \frac{%
5x_{1}+x_{2}}{2x_{1}}\right) \left(
b_{0}+b_{1}X+b_{2}X^{2}++b_{3}X^{3}\right) ~,  \notag \\
\dot{x}_{2} &=&x_{2}\left( a_{0}+a_{1}X+a_{2}X^{2}\right) -\left( \frac{%
4x_{1}-x_{2}}{x_{1}}\right) \left(
b_{0}+b_{1}X+b_{2}X^{2}++b_{3}X^{3}\right) ~,  \notag \\
X &=&2x_{1}+x_{2}~.  \label{A23}
\end{eqnarray}%
Note that the right hand sides of these ODEs (\ref{A23}) are \textit{%
polynomial} only if \textit{all} the coefficients $b_{\ell }$ vanish.

\textbf{Model A2.4}: $m_{1}=3,~m_{2}=1;$ $L=3;$ $\alpha _{0}=0,~\alpha
_{\ell }=\left( -1\right) ^{\ell -1}3c_{\ell },~\beta _{\ell }=\left(
-1\right) ^{\ell -1}c_{\ell -1},~\ell =1,2,3;$ $\gamma _{\ell }=0,~\ell
=0,1,2,3.$%
\begin{equation}
\dot{x}_{n}=x_{n}\left( c_{0}+c_{1}X+c_{2}X^{2}\right)
~,~~~n=1,2,~~~X=\left( x_{1}\right) ^{2}x_{2}~.  \label{A24}
\end{equation}

\textbf{Model A2.5}: $m_{1}=2,~m_{2}=3;$ $L=3;$ $\alpha _{0}=0,~\alpha
_{\ell }=2c_{\ell -1},~\beta _{\ell }=3c_{\ell -1},~\ell =1,2,3;$ $\gamma
_{\ell }=0,~\ell =0,1,2,3.$%
\begin{equation}
\dot{x}_{n}=x_{n}\left( c_{0}+c_{1}X+c_{2}X^{2}\right)
~,~~~n=1,2,~~~X=x_{1}\left( x_{1}+2x_{2}\right) ~.  \label{A25}
\end{equation}

\textbf{Model A2.6}: $m_{1}=3,~m_{2}=2;$ $L=3;$ $\alpha _{0}=0,~\alpha
_{\ell }=\left( -1\right) ^{\ell -1}3c_{\ell -1},~\beta _{\ell }=\left(
-1\right) ^{\ell -1}2c_{\ell -1},~\ell =1,2,3;$ $\gamma _{\ell }=0,~\ell
=0,1,2,3.$%
\begin{equation}
\dot{x}_{n}=x_{n}\left( c_{0}+c_{1}X+c_{2}X^{2}\right)
~,~~~n=1,2,~~~X=\left( x_{1}\right) ^{2}x_{2}~.  \label{A26}
\end{equation}%
\textbf{Remark 3.2-1}. Note that the systems of ODEs (\ref{A22}) and (\ref%
{A25}) are identical, and likewise the systems of ODEs (\ref{A24}) and (\ref%
{A26}) are identical. $\square $

\bigskip

\subsection{Models of type \textbf{A3}}

In the cases listed in this \textbf{Subsection} the variables $%
y_{m_{1}}\left( t\right) $ and $y_{m_{2}}\left( t\right) $ are supposed to
satisfy the system \textbf{A3} of $2$ coupled ODEs (see (\ref{A3}) and
\textbf{Appendix A}), with the indicated assignments of the various
parameters.

\textbf{Model A3.1}: $m_{1}=1,~m_{2}=2;$ $\alpha _{0}=-3a,$~$\alpha
_{1}=-9b;~\beta _{0}=-2a,~\beta _{1}=-2b;$%
\begin{eqnarray}
\dot{x}_{1} &=&a+b\left[ \left( x_{1}\right) ^{2}+7x_{1}x_{2}+\left(
x_{2}\right) ^{2}\right] ~,  \notag \\
\dot{x}_{2} &=&a+b\left[ 7\left( x_{1}\right) ^{2}+4x_{1}x_{2}-2\left(
x_{2}\right) ^{2}\right] ~.  \label{A31}
\end{eqnarray}

\textbf{Model A3.2}: $m_{1}=1,~m_{2}=3;$ $\alpha _{0}=-18a,$~$\alpha
_{1}=-486b;~\beta _{0}=-2a,~\beta _{1}=-2b;$%
\begin{eqnarray}
\dot{x}_{1} &=&\left( x_{1}\right) ^{-1}\left\{ a\left( 5x_{1}+x_{2}\right)
\right.  \notag \\
&&\left. +b\left[ 32\left( x_{1}\right) ^{4}-131\left( x_{1}\right)
^{3}x_{2}-51\left( x_{1}x_{2}\right) ^{2}-11x_{1}\left( x_{2}\right)
^{3}-\left( x_{2}\right) ^{4}\right] \right\} , \\
\dot{x}_{2} &=&\left( x_{1}\right) ^{-1}\left\{ 2a\left( 4x_{1}-x_{2}\right)
\right.  \notag \\
&&\left. -2b\left[ 32\left( x_{1}\right) ^{4}+112\left( x_{1}\right)
^{3}x_{2}-51\left( x_{1}x_{2}\right) ^{2}-11x_{1}\left( x_{2}\right)
^{3}-\left( x_{2}\right) ^{4}\right] \right\} .
\end{eqnarray}

\bigskip

\section{Outlook}

Space limitations do not allow to report here some natural developments of
the results reported above, such as the identification of the \textit{more
general solvable} systems which obtain from those reported in the preceding
\textbf{Section 3} via \textit{linear} transformations of the dependent
variables $x_{n}\left( t\right) $; the interested reader is referred to the
analogous developments discussed in\textbf{\ Section 5} of Ref. \cite%
{CP2019a}.

\bigskip

\section{Acknowledgements}

FP likes to thank the Physics Department of the University of Rome\ "La
Sapienza" for the hospitality from April 2018 to January 2019 (during her
sabbatical, when the results reported in this paper were obtained), and the
organizers of DICE2018 (where these results were reported), especially Prof.
Hans Thomas Elze.

\bigskip

\section{Appendix A}

The $2$ ODEs of the system \textbf{A1} (see (\ref{A1})) are uncoupled, and
each of them is clearly solvable by quadratures. Note in particular that the
initial-value problem of the ODE
\begin{subequations}
\begin{equation}
\dot{y}=ay+by^{m+1}
\end{equation}%
is given by the explicit formula%
\begin{equation}
y\left( t\right) =y\left( 0\right) \exp \left( at\right) \left\{ 1+\left(
b/a\right) \left[ y\left( 0\right) \right] ^{m}\left[ 1-\exp \left(
mat\right) \right] \right\} ^{-1}~.
\end{equation}

To solve the system \textbf{A2} one treats firstly the first of the $2$ ODEs
(\ref{A2})---essentially as just above. Then one notes---see \textbf{%
Appendix A }of Ref. \cite{CP2019b}---that the solution of the second ODE of
the system (\ref{A2}) reads
\end{subequations}
\begin{subequations}
\begin{equation}
y_{m_{2}}\left( t\right) =F\left( t\right) \left[ y_{m_{2}}\left( 0\right)
+\int_{0}^{t}dt^{\prime }~\left[ F\left( t^{\prime }\right) \right]
^{-1}\sum_{\ell =0}^{L}\left\{ \gamma _{\ell }\left[ y_{m_{1}}\left(
t^{\prime }\right) \right] ^{\ell -1+\left( m_{2}/m_{1}\right) }\right\} %
\right]
\end{equation}%
with%
\begin{equation}
F\left( t\right) =\exp \left\{ \int_{0}^{t}\left[ dt^{\prime }\sum_{\ell
=1}^{L}\left\{ \beta _{\ell }\left[ y_{m_{1}}\left( t^{\prime }\right) %
\right] ^{\ell -1}\right\} \right] \right\} ~.  \label{A2F}
\end{equation}

Explicit solutions can be easily obtained in the following cases:
\end{subequations}
\begin{subequations}
\label{mmm}
\begin{equation}
m_{1}=1~,~~~m_{2}=2,3~,~~~m=m_{2}/m_{1}~,
\end{equation}%
\begin{equation}
\dot{y}_{m_{1}}=\alpha _{0}+\alpha _{1}y_{m_{1}}+\alpha _{2}\left(
y_{m_{1}}\right) ^{2}~,
\end{equation}%
\begin{equation}
\dot{y}_{m_{2}}=y_{m_{2}}\left( \beta _{0}+\beta _{1}y_{m_{1}}\right)
+\gamma _{0}\left( y_{m_{1}}\right) ^{-1+m}+\gamma _{1}\left(
y_{m_{1}}\right) ^{m}+\gamma _{2}\left( y_{m_{1}}\right) ^{1+m}~;
\end{equation}%
\begin{eqnarray}
y_{_{m_{1}}}\left( t\right) &=&\frac{y_{_{m_{1}}}\left( 0\right) \left[
1+\left( \Delta /\alpha _{1}\right) \tanh \left( \Delta t\right) \right]
-2\left( \alpha _{0}/\alpha _{1}\right) \tanh \left( \Delta t\right) }{%
1-\left\{ \left[ 2\alpha _{2}y_{_{m_{1}}}\left( 0\right) +\Delta \right]
/\alpha _{1}\right\} \tanh \left( \Delta t\right) }~,  \notag \\
\Delta ^{2} &=&\left( \alpha _{1}\right) ^{2}-4\alpha _{0}\alpha _{2}~,
\end{eqnarray}%
\begin{equation}
y_{m_{2}}\left( t\right) =f\left( t\right) \left[ y_{m_{2}}\left( 0\right)
+\int_{0}^{t}dt^{\prime }~\left[ f\left( t^{\prime }\right) \right]
^{-1}\sum_{\ell =0}^{L}\left\{ \gamma _{\ell }\left[ y_{m_{1}}\left(
t^{\prime }\right) \right] ^{\ell -1+m}\right\} \right] ~,
\end{equation}%
\begin{equation}
f\left( t\right) =\exp \left\{ \int_{0}^{t}\left[ dt^{\prime }\sum_{\ell
=1}^{L}\left\{ \beta _{\ell }\left[ y_{\tilde{m}_{1}}\left( t^{\prime
}\right) \right] ^{\ell -1}\right\} \right] \right\} ~.
\end{equation}

Finally, to identify the solution of the system \textbf{A3} (see (\ref{A3}))
we note that time-differentiation of the first of its $2$ ODEs entails that $%
y_{m_{1}}\left( t\right) $ satisfies the decoupled second-order ODE
\end{subequations}
\begin{equation}
\ddot{y}_{m_{1}}=\alpha _{1}\left[ \beta _{0}\left( y_{m_{1}}\right)
^{-1+m}+\beta _{1}\left( y_{m_{1}}\right) ^{-1+2m}\right]
~,~~~m=m_{2}/m_{1}~;
\end{equation}%
hence (see (\ref{A3})) $y_{1}\left( t\right) $ is an elliptic (for $m=2$) or
hyperelliptic (for $m=3$) function; and likewise for $y_{2}\left( t\right) $
(see the first of the $2$ ODEs (\ref{A3})).


\bigskip

\begin{thebibliography}{99}
\bibitem{C1978} F. Calogero, "Motion of Poles and Zeros of Special Solutions
of Nonlinear and Linear Partial Differential Equations, and Related
"Solvable" Many Body Problems", Nuovo Cimento \textbf{43B}, 177-241 (1978).

\bibitem{C2001} F. Calogero, \textit{Classical many-body problems amenable
to exact treatments}, Lecture Notes in Physics Monograph \textbf{m66},
Springer, Heidelberg, 2001 (749 pages).

\bibitem{C2008} F. Calogero, \textit{Isochronous systems}, Oxford University
Press, Oxford, U. K., (2008); paperback (2012).

\bibitem{C2018a} F. Calogero, \textit{Zeros of Polynomials and Solvable
Nonlinear Evolution Equations}, Cambridge University Press, Cambridge, U.
K., 2018 (168 pages).

\bibitem{C2018b} F. Calogero, \textquotedblleft Systems of
nonlinearly-coupled differential equations solvable by algebraic
operations\textquotedblright , in \textit{Nonlinear Systems and Their
Remarkable Mathematical Structures }(edited by N. Euler), CRC Press (Taylor
\& Francis), 2018, pages 1-14.

\bibitem{GS2005} D. G\'{o}mez-Ullate and M. Sommacal, "Periods of the
Goldfish Many-Body Problem", J. Nonlinear Math. Phys. \textbf{12}, Suppl.
\textbf{1}, 351-362 (2005).

\bibitem{BC2018} O. Bihun and F. Calogero, \textquotedblleft Time-dependent
polynomials with \textit{one double} root, and related new solvable systems
of nonlinear evolution equations\textquotedblright , Qual. Theory Dyn. Syst.
(published online: 26 July 2018). doi.org/10.1007/s12346-018-0282-3;
http://arxiv.org/abs/1806.07502.

\bibitem{B2018} O. Bihun, "Time-dependent polynomials with one multiple root
and new solvable dynamical systems", arXiv:1808.00512v1 [math-ph] 1 Aug 2018.

\bibitem{CP2019a} F. Calogero and F. Payandeh, "Polynomials with multiple
zeros and solvable dynamical systems including models in the plane with
polynomial interactions", J. Math. Phys. (submitted).

\bibitem{CP2019b} F. Calogero and F. Payandeh, "Solvable dynamical systems
in the plane with polynomial interactions", to be published as a chapter in
a collective book to celebrate the 65th birthdate of Emma Previato (in
press).
\end{thebibliography}
\end{document}